\pdfoutput=1
\documentclass[12pt]{article}
\usepackage[utf8]{inputenc}
\usepackage{amsmath}
\usepackage{mathtools}
\usepackage{booktabs}
\usepackage{cite}
\usepackage{amssymb}
\usepackage[titletoc,title]{appendix}
\usepackage{mathrsfs}
\usepackage{float}
\usepackage{multirow}
\usepackage{graphicx,caption,subcaption}
\usepackage[space]{grffile}
\usepackage{fullpage}
\usepackage{color}
\usepackage{soul}
\usepackage[dvipsnames]{xcolor}

\definecolor{darkblue}{rgb}{0.1,0.1,.7}
\usepackage[breaklinks=true,linktocpage=true,
colorlinks=true,urlcolor=darkblue,linkcolor=darkblue,
citecolor=darkblue,pdfpagelabels=true,hypertexnames=true,
plainpages=false,naturalnames=false,]{hyperref}

\usepackage{datetime}
\newdateformat{monthyeardate}{%
  \monthname[\THEMONTH], \THEYEAR}
\date{\monthyeardate\today}

\newcommand{\overbar}[1]{\mkern1.5mu\overline{\mkern-1.5mu#1\mkern-1.5mu}\mkern 1.5mu}

\def\simlt{\mathrel{\lower2.5pt\vbox{\lineskip=0pt\baselineskip=0pt
           \hbox{$<$}\hbox{$\sim$}}}}
\def\simgt{\mathrel{\lower2.5pt\vbox{\lineskip=0pt\baselineskip=0pt
           \hbox{$>$}\hbox{$\sim$}}}}

\begin{document}

\renewcommand{\arraystretch}{1.3}
\thispagestyle{empty}

{\hbox to\hsize{\vbox{\noindent November 2021}}}

\noindent
\vskip2.0cm
\begin{center}

{\Large\bf Reheating after inflation by supersymmetry breaking}

\vglue.3in

\hspace{-2pt}Yermek Aldabergenov,${}^{a,b,}$\footnote{yermek.a@chula.ac.th} Ignatios Antoniadis,${}^{c,}$\footnote{antoniad@lpthe.jussieu.fr} Auttakit Chatrabhuti,${}^{a,}$\footnote{auttakit.c@chula.ac.th} Hiroshi Isono${}^{a,}$\footnote{hiroshi.isono81@gmail.com}
\vglue.1in

${}^a$~{\it Department of Physics, Faculty of Science, Chulalongkorn University,\\ Phayathai Road, Pathumwan, Bangkok 10330, Thailand}\\
${}^b$~{\it Department of Theoretical and Nuclear Physics, Faculty of Physics and Technology,\\
Al-Farabi Kazakh National University, 71 Al-Farabi Ave., Almaty 050040, Kazakhstan}\\
${}^c$~{\it Laboratoire de Physique Th\'eorique et Hautes Energies (LPTHE), Sorbonne Universit\'e,\\ CNRS, 4 Place Jussieu, 75005 Paris, France}
\vglue.1in

\end{center}

\vglue.3in

\begin{center}
\begin{abstract}
We study reheating after the end of inflation in models where the inflaton is the superpartner of goldstino and is charged under a gauged $U(1)$ R-symmetry. We consider two classes of models -- one is small field characterized by an almost flat K\"ahler space, and the other large field characterized by a hyperbolic K\"ahler space $SU(1,1)/U(1)$, while in both cases the inflaton superpotential is linear due to the R-symmetry. The inflationary observables of our models fit within 2$\sigma$ CMB values. Upon coupling the inflaton sector to the (supersymmetric) Standard Model, we compute the MSSM parameters, mass spectrum, and decay modes of the inflaton, with the resulting reheating temperature around $10^8$ GeV. We also find that both models can accommodate superheavy LSP dark matter, depending on the parameter choice.
\end{abstract}
\end{center}

\newpage

\tableofcontents

\setcounter{footnote}{0}

\section{Introduction}
In past works, a framework of natural inflation within supergravity was proposed, dubbed `inflation by supersymmetry breaking'~\cite{Antoniadis:2017gjr, Antoniadis:2019dpm}. The main idea is to identify the inflaton with the superpartner of the goldstino, in the presence of a gauged R-symmetry. The minimal field content consists of the inflaton chiral superfield $Z$ charged under an abelian Maxwell multiplet gauging the R-symmetry. The superpotential is then fixed by symmetry to be linear in $Z$, spontaneously breaking supersymmetry and in general R-symmetry with the gauge boson becoming massive by absorbing the phase of the inflaton. On the other hand, the K\"ahler potential can be expanded around the origin $Z=0$ which corresponds to a maximum of the scalar potential, where R-symmetry is restored. In the limit of vanishing gauge coupling, at the lowest order (canonical K\"ahler potential) the slow-roll parameter $\eta$ vanishes, while its first correction determines $\eta$. One therefore obtains a natural small-field hilltop type inflation around the maximum, ending as the inflaton rolls down towards the minimum. The latter is controlled by the second order correction and the D-term contribution which can tune the vacuum energy to zero, or a tiny positive value corresponding to the dark energy of the observable universe.

In this work, we couple the inflaton and supersymmetry breaking sector described above with the (supersymmetric) Standard Model (MSSM) and study the reheating after the end of inflation. In particular, we compute the superparticle spectrum~\cite{Antoniadis:2015mna}, the decay modes of the inflaton and the resulting reheating temperature. In principle, there are two distinct possibilities for the MSSM superpotential $\Omega$: (1) to be neutral under the R-symmetry, in which case the full superpotential is proportional to the goldstino superfield $Z$ and $\Omega$ is added to the goldstino decay constant; (2) to have the same R-charge as $Z$ so that Standard Model particles are neutral while their superpartners are charged, in which case $\Omega$ is just added to the supersymmetry breaking sector superpotential. Note that in this case the gauged R-symmetry contains the usual R-symmetry of the MSSM~\cite{Antoniadis:2015adn}. It turns out that both possibilities lead to similar results and thus we choose to perform the explicit analysis for case (1) and then (in the concluding section) comment on the corresponding changes for case (2). Of course, one could have more general situations with different R-charges that can be studied by extending our analysis in a straightforward way.

The outline of our paper is the following. In Section 2, we review briefly the framework of inflation by supersymmetry breaking and describe the coupling of the supersymmetry breaking sector to the MSSM. In Section 3, we specialise to a K\"ahler potential perturbatively expanded around the canonically flat case, compute the superparticle spectrum (soft scalar masses, gaugino masses and trilinear couplings), and calculate the decay modes of the inflaton and the resulting reheating temperature. In Section 4, we repeat the analysis for a non-flat hyperbolic K\"ahler potential. In Section 5, we consider the effects introducing $Z$-dependent wavefunctions for the matter fields in the K\"ahler potential, and derive the requirements for vacuum stability. In Section 6, we discuss Dark Matter possibilities within our model. Finally in Section 7, we present our conclusions, while in the Appendix we give the supergravity Lagrangian that we use throughout the paper.

\section{General setup}

The starting point is a class of models with gauged $U(1)_R$ phase symmetry, defined by K\"ahler potential and superpotential ($\kappa=M_P^{-1}$ is the gravitational constant),
\begin{gather}
	K(Z,\overbar{Z},\phi,\overbar{\phi})=\sum\phi\overbar{\phi}+J(Z\overbar{Z})~,\label{K_master}\\
	W(Z,\phi)=[a\kappa^{-3}+\Omega(\phi)]\kappa Z~,\label{W_master}
\end{gather}
where $Z$ is the inflaton/sgoldstino superfield, $\phi$ collectively denotes matter superfields, and $J$ is the inflaton K\"ahler potential. In the superpotential, $a$ is a dimensionless real constant, while $\Omega$ is the MSSM part,
\begin{equation}
	\Omega=\hat{y}_u\bar{u}Q H_u-\hat{y}_d\bar{d}Q H_d-\hat{y}_e\bar{e}L H_d+\hat{\mu}H_uH_d~.\label{Omega_master}
\end{equation}
Here $\bar u,\bar d,\bar e,Q,L,H_u,H_d$ are chiral superfields. As usual, we denote the corresponding SM matter fields (quarks, leptons, and Higgs fields) with the same character, while tildes will be used for their superpartners (squarks, sleptons, and Higgsinos). The un-normalized Yukawa couplings $y$ and the $\mu$-parameter are denoted by hats which will be removed after proper rescaling, once $Z$ settles at the minimum.

The total gauge group of the model is,
\begin{equation}
	SU(3)_c\times SU(2)_L\times U(1)_Y\times U(1)_R~.
\end{equation}
Squarks, sleptons, and Higgs scalars are neutral under $U(1)_R$, while $Z$ carries the same $R$-charge as the superpotential. The $R$-charges of the MSSM fermions will be fixed later.

The scalar potential is the sum $V=V_F+V_D$ where~\footnote{The mass dimensions of K\"ahler potential, superpotential, gauge kinetic function, Killing potential, and Killing vector are respectively,
\begin{equation}
	[K]=M^2~,~~~[W]=M^3~,~~~[f_{AB}]=M^0~,~~~[{\cal D}_{A}]=M^2~,~~~[X^I_{A}]=M~,\nonumber
\end{equation}
while scalar fields have canonical mass dimension $M$.}
\begin{gather}
	V_F=e^{\kappa^2K}\left\{ K^{I\bar{J}}D_IW D_{\bar{J}}\overbar{W}-3\kappa^2|W|^2\right\}~,\\
	V_D=\frac{1}{2}{\rm Re}(f^{AB}){\cal D}_{A}{\cal D}_{B}~.
\end{gather}
Here the indices $I,J$ run through all the chiral (super)fields, while $A,B$ are the gauge group indices. The relevant part of supergravity Lagrangian that we use here can be found in the Appendix and its derivation in Ref. \cite{Wess:1992cp}.

For the gauge kinetic matrix the following notation is used, $f^{AB}\equiv f_{AB}^{-1}$. K\"ahler covariant derivatives are defined as $D_I W\equiv W_I+\kappa^2K_I W$, where the indices denote the respective partial derivatives. The Killing potential and Killing vector are related by
\begin{equation}
	{\cal D}_{A}=iX_{A}^I\left(K_I+\kappa^{-2}\frac{W_I}{W}\right)~,
\end{equation}
where the gauge couplings and charges are included in the Killing vectors $X^I_{A}$. For example, if $Z$ transforms under $U(1)_R$ as $Z\rightarrow Ze^{-igq\vartheta}$ (where $\vartheta$ is a transformation parameter and $q$ is its $R$-charge), its Killing vector is $X^Z_R=-igqZ$. The gauge couplings of $U(1)_R$, $U(1)_Y$, $SU(2)_L$, and $SU(3)_c$ are $g$, $g_1$, $g_2$, and $g_3$, respectively.

We use a convention where the superpotential transforms under $U(1)_R$ with unit $R$-charge, $W\rightarrow We^{-ig\vartheta}$, and the fermionic superspace coordinate transforms with half-unit $R$-charge, $\theta\rightarrow\theta e^{-ig\vartheta/2}$. Then $Z$ has unit $R$-charge, while its fermionic partner has half-unit $R$-charge. In general, for a scalar field with $R$-charge $q$, its fermionic partner has $R$-charge $q-1/2$. With this convention let us write down in Table \ref{tab_fermi_charges} the fermion charges under the total gauge group of our model.

\begin{table}[hbt!]
\centering
\begin{tabular}{l r r r r r r r r r r r r}
\toprule
 & $Q$ & $\bar u$ & $\bar d$ & $L$ & $\bar e$ & $\tilde H_u$ & $\tilde H_d$ & $\zeta$ & $\lambda_R$ & $\lambda_1$ & $\lambda_2$ & $\lambda_3$ \\
\hline
$U(1)_R$ & $-\tfrac{1}{2}$ & $-\tfrac{1}{2}$ & $-\tfrac{1}{2}$ & $-\tfrac{1}{2}$ & $-\tfrac{1}{2}$ & $-\tfrac{1}{2}$ & $-\tfrac{1}{2}$ & $\tfrac{1}{2}$ & $\tfrac{1}{2}$ & $\tfrac{1}{2}$ & $\tfrac{1}{2}$ & $\tfrac{1}{2}$ \\
$U(1)_Y$ & $\tfrac{1}{6}$ & $-\tfrac{2}{3}$ & $\tfrac{1}{3}$ & $-\tfrac{1}{2}$ & $1$ & $\tfrac{1}{2}$ & $-\tfrac{1}{2}$ & $0$ & $0$ & $0$ & $0$ & $0$ \\
$SU(2)_L$ & $\bf 2$ & $\bf 1$ & $\bf 1$ & $\bf 2$ & $\bf 1$ & $\bf 2$ & $\bf 2$ & $\bf 1$ & $\bf 1$ & $\bf 1$ & $\bf 3$ & $\bf 1$ \\
$SU(3)_c$ & $\bf 3$ & $\bar{\bf 3}$ & $\bar{\bf 3}$ & $\bf 1$ & $\bf 1$ & $\bf 1$ & $\bf 1$ & $\bf 1$ & $\bf 1$ & $\bf 1$ & $\bf 1$ & $\bf 8$ \\\bottomrule
\hline
\end{tabular}
\captionsetup{width=.9\linewidth}
\caption{MSSM and $U(1)_R$ charges of the fermions. $\zeta$ is the inflatino, $\lambda_R$ is the $U(1)_R$ gaugino, and $\lambda_{1,2,3}$ are bino, wino, and gluino, respectively. The gravitino has the same $R$-charge as $\lambda_R$.}
\label{tab_fermi_charges}
\end{table}

In this class of models, the $Z$-dependent part of the potential drives inflation, after which $Z$ and its auxiliary field $F^Z$ settle at non-zero vacuum expectation values (VEVs), spontaneously breaking both supersymmetry (SUSY) and $U(1)_R$. At the vacuum, the gravitino mass and the the auxiliary fields of $Z$ and $U(1)_R$ are given by,
\begin{align}
\begin{split}
	m_{3/2} &=a\langle e^{\kappa^2J/2}|Z|\rangle~,\\
	\langle F^Z\rangle &=-a\langle e^{\kappa^2J/2}J^{Z\bar Z}(\kappa^{-2}+J_ZZ)\rangle~,\\
	\langle {\cal D}_R\rangle &=g\langle \kappa^{-2}+J_ZZ\rangle~,
\end{split}
\end{align}
where we assume that matter fields $\phi$ vanish at the minimum.

The Yukawa couplings $\hat{y}$ and the parameter $\hat{\mu}$ in \eqref{Omega_master} are related to their properly normalized versions as
\begin{equation}
	\{\hat{y},\hat\mu\}=\left\langle\frac{e^{-\kappa^2J/2}}{\kappa|Z|}\right\rangle\times \{y,\mu\}~.
\end{equation}
This is due to the overall factor of $e^K$ in the $F$-term potential, as well as the coupling of $\Omega$ to $Z$ as shown in Eq.~\eqref{W_master}. At the vacuum, $Z$ and $J(Z,\overbar Z)$ take non-vanishing VEVs, which leads to this rescaling.

As for the inflaton part $J$ of the K\"ahler potential, we consider two examples described below.

\section{Model I: (almost) flat K\"ahler space}

Since we require that matter fields vanish at the minimum, vacuum structure is defined entirely by the choice of the K\"ahler potential $J(Z,\overbar Z)$ (as the superpotential is already fixed). One example of a suitable (for inflation and SUSY breaking) almost canonical K\"ahler potential was given in Ref.~\cite{Antoniadis:2017gjr}, which uses a non-perturbative correction of the form $\sim Z\overbar Ze^{\kappa^2Z\overbar Z}$. Here we would like to introduce a simpler choice of $J$ with finite number of perturbative corrections, namely,
\begin{equation}
    J=Z\overbar Z+\alpha\kappa^2(Z\overbar Z)^2+\beta\kappa^4(Z\overbar Z)^3~,\label{J_1}
\end{equation}
where the parameters $\alpha$ and $\beta$ are dimensionless. One should think of the above form as a perturbative expansion around the canonical kinetic terms with coefficients less than unity.

Let us study the vacuum and the possibility of inflation in this model by ignoring matter fields so that $K=J$ is given by \eqref{J_1}, and $W=\kappa^{-2}aZ$. Then the scalar potential reads,
\begin{align}
    \kappa^{-4}V=a^2\exp\left(|\kappa Z|^2+\alpha|\kappa Z|^4+\beta|\kappa Z|^6\right)&\left\{\frac{(1+|\kappa Z|^2+2\alpha|\kappa Z|^4+3\beta|\kappa Z|^6)^2}{1+4\alpha|\kappa Z|^2+9\beta|\kappa Z|^4}-3|\kappa Z|^2\right\}\nonumber\\
    &+\frac{g^2}{2}\left(1+|\kappa Z|^2+2\alpha|\kappa Z|^4+3\beta|\kappa Z|^6\right)^2~,\label{model_I_potential}
\end{align}
where we set gauge kinetic function $f=1$ for now.

First, consider the limit of vanishing $U(1)_R$ coupling $g$, in which case the parameter of the superpotential $a$ becomes an overall factor of the scalar potential. In this setup it is possible to obtain a ``double-well" potential with local maximum at $|Z|=0$ and Minkowski minimum away from $|Z|=0$.

When $\alpha=0$ (and $g=0$) the Minkowski vacuum equations $V=\partial_ZV=0$ can be solved exactly, which yields
\begin{equation}
    \beta=4/81~,~~~|\langle\kappa Z\rangle|=\sqrt{3/2}~.
\end{equation}
However in this case the potential is convex ($V_{Z\bar Z}>0$) around $Z=0$ and slow-roll hilltop inflation is not possible. The problem is solved if we add small $\alpha$. One example of a suitable scalar potential is given by the parameter choice
\begin{equation}
    \alpha=0.005~,~~~\beta=0.0494~,\label{param_I_g=0}
\end{equation}
with the inflaton VEV $|\langle\kappa Z\rangle|=1.22$.\footnote{Although $\langle\kappa Z\rangle$ is bigger than unity, the expansion in \eqref{J_1} is still valid taking into account the small coefficients.} This gives rise to inflation with
\begin{equation}
    n_s=0.9615~,~~~r=8.85\times 10^{-5}~,~~~H_{\rm inf}=2.33\times 10^{12}~{\rm GeV}~,
\end{equation}
for $60$ e-folds. The scalar amplitude of $A_s=2.1\times 10^{-9}$ \cite{Akrami:2018odb} can be used to fix the parameter $a$ at $1.66\times 10^{-6}$, which also controls the gravitino mass. Thus, the SUSY breaking scale (after inflation) is close to the inflationary scale in this class of models.

Although the global $U(1)_R$ ($g=0$) case is viable for inflation, it includes a massless $R$-axion, which motivates us to gauge the $R$-symmetry so that the axion is absorbed by the massive $U(1)_R$ vector, when the $R$-symmetry is spontaneously broken (an alternative would be to give mass to the axion by explicit $R$-symmetry breaking terms). In this case the gauge kinetic functions will be fixed by the requirement of anomaly cancellation via Green--Schwarz mechanism.

When the gauge coupling $g$ is turned on, the requirement of viable inflation puts an upper bound on it of around $g\sim 0.8a$ \footnote{Concretely, the bound avoids too small $n_s$.} (no lower bound). This freedom to choose $g$ can be used to control soft scalar masses to some extent. In Subsection \ref{Subsec_par_space} we will explore the allowed parameter space of this model in more detail.

\subsection{Soft scalar masses}\label{subsec_m0}
	
In our model defined by \eqref{K_master} and \eqref{W_master} (with general $J$) soft scalar masses are universal,
\begin{equation}
	m_Q^2=m_u^2=m_d^2=m_L^2=m_e^2=m_{H_u}^2=m_{H_d}^2=m^2_0~,
\end{equation}
where $m_0^2$ is given by
\begin{equation}
	m_0^2=\kappa^2\langle J_{Z\bar Z}F^Z\overbar{F}^Z\rangle-2m_{3/2}^2~,
\end{equation}
and for the MSSM $\mu$-parameter we assume $|\mu|\ll |m_0|$ to avoid extreme fine-tuning of the Higgs boson mass (since $m_0$ is close to the inflationary scale). From the requirement of (near-)Minkowski minimum, we have the relation,~\footnote{Here the $U(1)_R$ gauge kinetic function is approximated by $\langle f_{R}\rangle\approx 1$, as will be justified below (see Eq.~\eqref{f_R_approx}).}
\begin{equation}
	\langle V\rangle=\langle J_{Z\bar Z}F^Z\overbar{F}^Z\rangle-3\kappa^{-2}m_{3/2}^2+\tfrac{1}{2}\langle{\cal D}_R\rangle^2=0~.\label{V_vac}
\end{equation}
The relation \eqref{V_vac} allows us to rewrite $m_0^2$ in terms of the $D$-term contribution,
\begin{equation}
	m_0^2=m_{3/2}^2-\tfrac{\kappa^2}{2}\langle{\cal D}_R\rangle^2~,\label{m_0_expression}
\end{equation}
and this leads to the requirement $m_{3/2}>\kappa\langle{\cal D}_R\rangle/\sqrt{2}$, in order to avoid tachyonic instabilities in the MSSM sector.

For the bilinear $H_u H_d$ coupling we have
\begin{equation}
	e^{-1}{\cal L}\supset -B_0\mu H_uH_d+{\rm h.c.}~,
\end{equation}
where
\begin{equation}
	B_0=\frac{\kappa^2\langle J_{Z\bar Z}F^Z\overbar F^Z\rangle-m^2_{3/2}}{m_{3/2}}~.\label{B_0_def}
\end{equation}

\subsection{Exploring the parameter space}\label{Subsec_par_space}

In this subsection we analyze the parameter space of the model, namely the allowed values of the parameters $\alpha$ and $\beta$ of the K\"ahler potential.

\begin{figure}
\centering
  \centering
  \includegraphics[width=.65\linewidth]{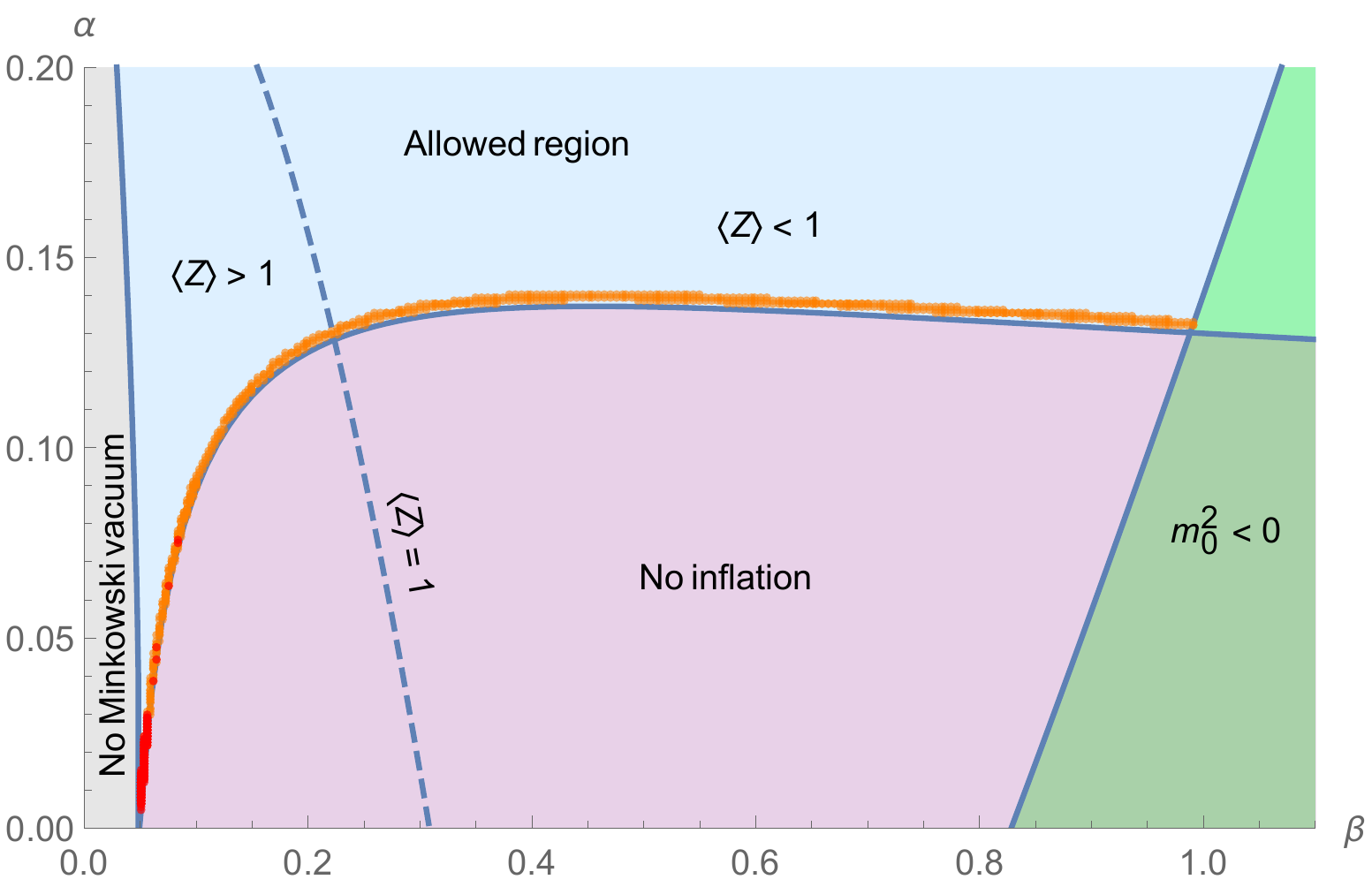}
\captionsetup{width=1\linewidth}
\caption{Constraints on the parameters $\alpha$ and $\beta$ of the K\"ahler potential of model I. Thin orange and red regions represent $n_s>0.954$ and $n_s>0.9607$. $\langle Z\rangle$ is in Planck units. Although not shown in the plot, the effective parameter $g/a$ is fixed at each point in the plot by the requirements (\textit{i})--(\textit{iii}), taking the values from $0$ to $\sim 0.75$.}\label{Fig_par_space}
\end{figure}

We constrain the parameter space $\alpha-\beta$ by requiring (\textit{i}) stable Minkowski vacuum away from $Z=0$, (\textit{ii}) concave potential ($V_{Z\bar Z}<0$) around the origin for slow-roll hilltop inflation, and (\textit{iii}) non-tachyonic soft scalar mass, $m_0^2>0$ (see Eq.~\eqref{m_0_expression}). The parameter $a$ can be factored out from the total scalar potential (thus determined by the scale of inflation), while the effective gauge parameter $g/a$ is used as a variable in order to satisfy the requirements (\textit{i})--(\textit{iii}). The allowed parameter values are shown in Figure \ref{Fig_par_space} as the blue shaded region.\,\footnote{We also checked negative values of $\alpha$ and $\beta$ which do not seem to allow for a suitable scalar potential, although the situation may change if further corrections to the K\"ahler potential are added.} In this allowed region we also compute the values of the spectral index $n_s$: the orange dots represent $n_s>0.954$ (around $2\sigma$ values from CMB), and the red dots on the left-side represent $n_s>0.9607$ ($1\sigma$ values). The effective parameter $g/a$ varies from $g/a=0$ (the line between blue and gray regions on the left) to $g/a\approx 0.75$ in the allowed region.

\begin{figure}
\centering
  \centering
  \includegraphics[width=.65\linewidth]{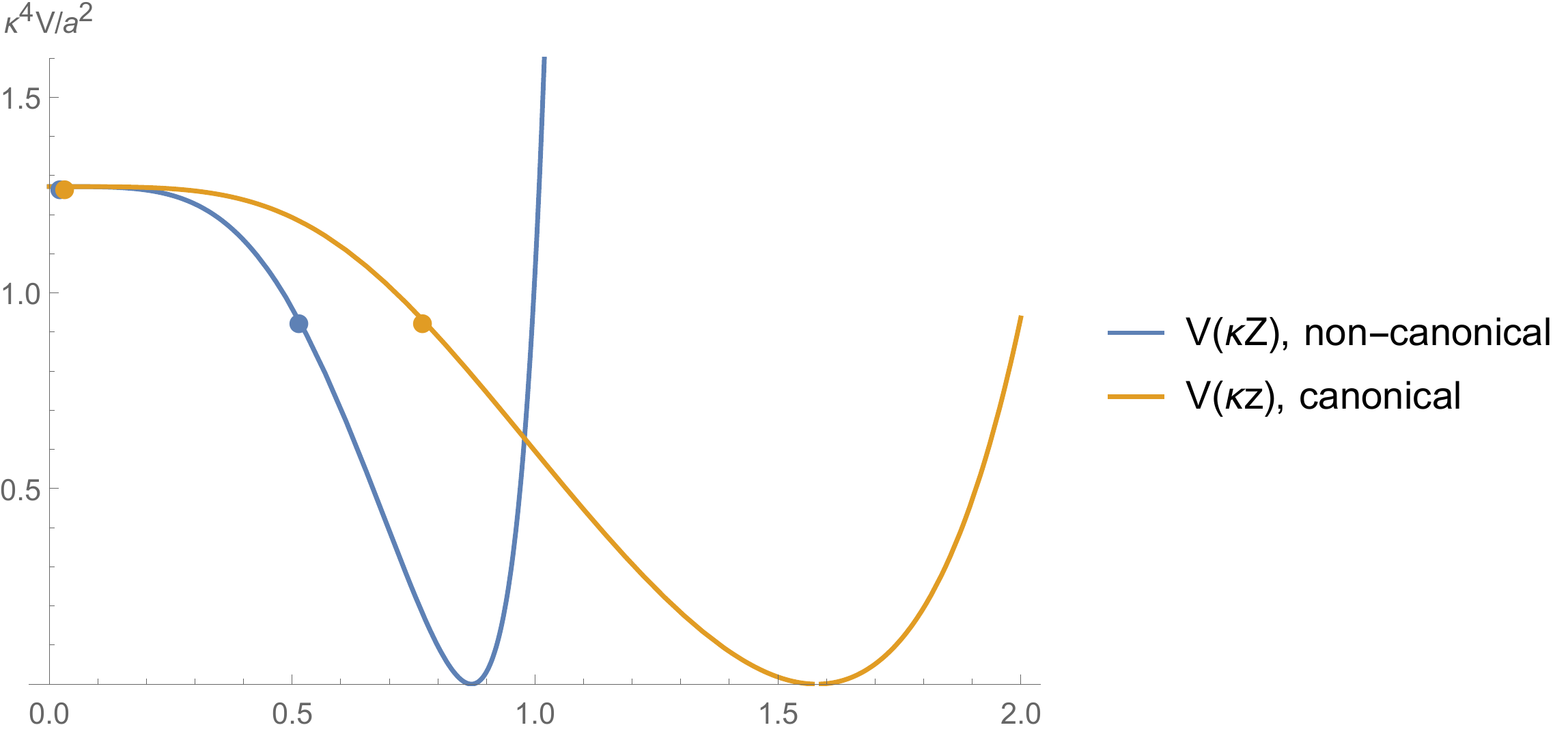}
\captionsetup{width=1\linewidth}
\caption{Scalar potential \eqref{model_I_potential} of model I for the parameter set \eqref{model_I_parameters}. Both non-canonical ($Z$) and canonical ($z$) parametrizations are shown, where the latter is found numerically. The markers represent the start and end of $60$ e-folds of inflation (the starting point of inflation almost coincides for the two curves).}\label{Fig_V_1}
\end{figure}

As a concrete example we choose the following parameter values
\begin{equation}
    \alpha=0.139~,~~~\beta=0.6~,~~~g/a=0.7371~,~~~a=2.05\times 10^{-7}~,\label{model_I_parameters}
\end{equation}
which leads to the inflationary parameters
\begin{equation}
    n_s=0.9543~,~~~r=1.72\times 10^{-6}~,~~~H_{\rm inf}=3.25\times 10^{11}~{\rm GeV}~,
\end{equation}
and the scalar potential depicted in Figure \ref{Fig_V_1} where the non-canonical scalar $Z$ is shown in blue, while the canonically normalized scalar $z$, found numerically, is shown in orange. The corresponding inflaton VEV is $\langle\kappa Z\rangle=0.89$. We find that in the allowed parameter region, the smallest possible value of $\langle\kappa Z\rangle$ compatible with a $2\sigma$ CMB constraint on $n_s$ is around $0.8$, while $1\sigma$ values of $n_s$ favour $\langle\kappa Z\rangle>1$.

\subsection{Gaugino masses}

The MSSM gaugino masses are generated at one loop via the Green--Schwarz mechanism of anomaly cancellation, where the gauge anomalies due to triangle diagrams involving the fermions (all the fermions of the model carry non-zero $R$-charges) are cancelled by appropriate $U(1)_R$ transformations of the following terms depending on the imaginary part of the gauge kinetic matrix,
\begin{equation}
	e^{-1}{\cal L}\supset\tfrac{1}{8}{\rm Im}(f_{AB})\epsilon^{mnkl}F^{A}_{mn}F^{B}_{kl}~.
\end{equation}

The gauge kinetic matrix of the model has the form,
\begin{equation}
	f_{AB}=
	\begin{pmatrix}
	f_R & & & \\
	& f_1 & & \\
	& & f_2 & \\
	& & & f_3
	\end{pmatrix}~,
\end{equation}
where $f_{R,1,2,3}$ are gauge kinetic functions for $U(1)_R$, $U(1)_Y$, $SU(2)_L$, and $SU(3)_c$, respectively. To cancel the anomalies we fix these kinetic functions as,
\begin{align}
	f_R &=1+\beta_R\log({\kappa Z})~,\label{f_R_def}\\
	f_{a} &=1+\beta_{a}\log({\kappa Z})~,\label{f_a_def}
\end{align}
where $a=1,2,3$ stands for the Standard Model gauge groups. Here $\beta$ are constants which we determine by using the methods described in Refs. \cite{Freedman:2005up,Elvang:2006jk,Antoniadis:2014iea}. The result is,
\begin{equation}
\beta_R=-\frac{g^2}{3\pi^2}~,~~~\beta_1=-\frac{11g_1^2}{8\pi^2}~,~~~\beta_2=-\frac{5g_2^2}{8\pi^2}~,~~~\beta_3=-\frac{3g_3^2}{8\pi^2}~,\label{beta_values}
\end{equation}
where $\beta_R$ is found from the cancellation of $U(1)_R^3$ anomaly, $\beta_1$ from $U(1)_R\times U(1)_Y^2$ anomaly, $\beta_2$ from $U(1)_R\times [SU(2)_L]^2$ anomaly, and $\beta_3$ from $U(1)_R\times [SU(3)_c]^2$ anomaly.

The values of $\beta_{a}$ are the same as in the model of Ref. \cite{Antoniadis:2015mna}, because the MSSM fermions in the two models have the same $R$-charges, while $\beta_R$ is different due to the difference in the hidden sector fermion (inflatino) $R$-charges. Since $g/\kappa$ in our models is not far from the Hubble scale (e.g. model I with the parameter choice \eqref{model_I_parameters} leads to $g\sim 10^{-7}$), we have
\begin{equation}
	f_R=1+\beta_R\log(\kappa Z)\approx 1~,\label{f_R_approx}
\end{equation}
if $\kappa Z$ is around unity.
Gauged $U(1)_R$ also leads to a gravitational anomaly which can be cancelled in a similar fashion (see for example Refs. \cite{Freedman:2005up,Elvang:2006jk,Antoniadis:2014iea}).

This brings us to the MSSM gaugino masses,
\begin{align}
	m_{ab}=\frac{1}{2}\left|\langle F^Z\partial_Z f_{ab}\rangle\right|=\frac{a}{2}\left|\left\langle e^{\kappa^2J/2}J^{Z\bar Z}(\kappa^{-2}+J_{\bar Z}\overbar Z)\partial_Z f_{ab} \right\rangle\right|~.
\end{align}
Using Eq. \eqref{f_a_def} we get,
\begin{equation}
	m_{a}=\left|\frac{\langle\kappa F^Z\rangle\beta_{a}}{2\langle\kappa Z\rangle}\right|~,\label{m_a_def}
\end{equation}
where we denote $m_{a}\equiv m_{aa}$. Finally, $m_{a}$ should be rescaled after taking into account non-canonical kinetic terms of the gaugini,
\begin{equation}
	e^{-1}{\cal L}\supset -\tfrac{i}{2}\langle{\rm Re}f_{a}\rangle\lambda^{a}\sigma^mD_m\bar\lambda^{a}+{\rm h.c.}=-\tfrac{i}{2}(1+\beta_{a}\log\langle\kappa Z\rangle)\lambda^{a}\sigma^mD_m\bar\lambda^{a}+{\rm h.c.}
\end{equation}
However, if $|\beta_a|\log\langle\kappa Z\rangle\ll 1$, as in the models that we consider here, the rescaling of the gaugini can be neglected.

\subsection{Trilinear couplings}

The trilinear couplings between the MSSM scalars are
\begin{equation}
	e^{-1}{\cal L}\supset -A_0(y_u\bar{\tilde u}\tilde QH_u-y_d\bar{\tilde d}\tilde QH_d-y_e\bar{\tilde e}\tilde LH_d)-\mu(y_u\bar{\tilde u}\tilde Q\overbar H_d-y_d\bar{\tilde d}\tilde Q\overbar H_u-y_e\bar{\tilde e}\tilde L\overbar H_u)+{\rm h.c.}~,
\end{equation}
where for $A_0$ we have
\begin{equation}
	A_0=\frac{\kappa^2\langle J_{Z\bar Z}F^Z\overbar F^Z\rangle}{m_{3/2}}~,
\end{equation}
which is related to $B_0$ from Eq.~\eqref{B_0_def} as $A_0=B_0+m_{3/2}$ (similarly to the model of Ref. \cite{Antoniadis:2015mna}).

\subsection{Soft parameters and mass spectrum}

Here we show explicit values of the MSSM soft parameters for the parameter set \eqref{model_I_parameters}, as well as the mass spectrum of the model. The results are summarized in Table \ref{tab_soft_I}, where we take one-loop values of the Standard Model gauge couplings~\footnote{We choose non-SUSY running of the couplings because SUSY breaking scale is very high in our models.} at the reheating temperature $10^{8}$ GeV (estimated in the next subsection),
\begin{equation}
    g_1=0.5~,~~~g_2=0.59~,~~~g_3=0.72~.
\end{equation}
As for the $U(1)_R$ gauge boson, its mass generated by the Higgs mechanism is $9.61\times 10^{11}$ GeV, close to the inflaton mass.

The parameters $A_0$ and $B_0$ are estimated as
\begin{equation}
    A_0=1.6\times 10^{12}~{\rm GeV}~,~~~B_0=8.46\times 10^{11}~{\rm GeV}~.
\end{equation}

\begin{table}[hbt!]
\centering
\begin{tabular}{r r r r r r r}
\toprule
$m_z$ & $m_\zeta$ & $m_{3/2}$ & $m_0$ & $m_1$ & $m_2$ & $m_3$ \\
\hline
$1.25\times 10^{12}$ & $6.15\times 10^{11}$ & $7.51\times 10^{11}$ & $2.68\times 10^{11}$ & $1.03\times 10^{10}$ & $6.54\times 10^{9}$ & $5.84\times 10^{9}$ \\\bottomrule
\hline
\end{tabular}
\captionsetup{width=1\linewidth}
\caption{Masses (in GeV) of inflaton, inflatino, gravitino, and MSSM sparticles derived from model I with parameter set \eqref{model_I_parameters}.}
\label{tab_soft_I}
\end{table}

\begin{figure}
\centering
  \centering
  \includegraphics[width=.55\linewidth]{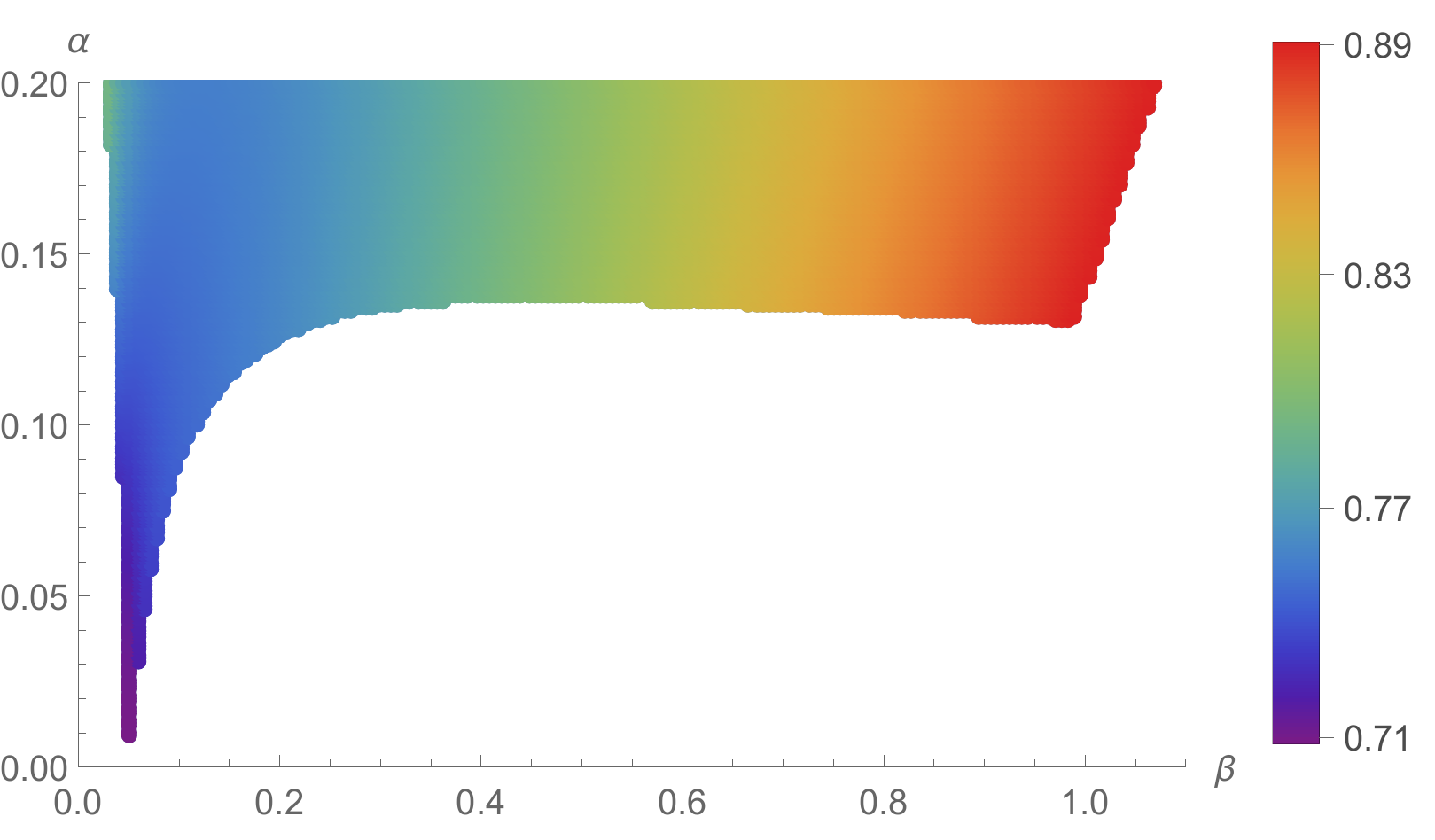}
\captionsetup{width=1\linewidth}
\caption{The color-coded ratio $m_z/(2m_{3/2})$ in the allowed region of $\alpha$--$\beta$ plane.}\label{Fig_mr}
\end{figure}

As can be seen from Table \ref{tab_soft_I}, the inflaton mass is smaller than two times the gravitino mass, $m_z<2m_{3/2}$, which prohibits the perturbative decay of the inflaton into gravitini. In order to see if this holds for other parameter values, we draw the ratio $m_z/(2m_{3/2})$ as a function of $\alpha$ and $\beta$ in Figure \ref{Fig_mr}, showing that $m_z$ is smaller than $2m_{3/2}$ in the relevant parameter range; due to the observed value of the spectral index, the relevant parameter region is constrained to a narrow strip around the lower limit of the allowed space, as shown in Figure \ref{Fig_par_space}. Therefore the inflaton-gravitino direct decay is forbidden.\footnote{Other mechanisms of gravitino production, such as non-perturbative production during preheating \cite{Kallosh:1999jj,Addazi:2016bus,Hasegawa:2017hgd} need further investigation.}

\subsection{Inflaton-MSSM interactions}\label{Subsec_Inflaton-MSSM}

Here we derive trilinear and quartic interactions between the inflaton and MSSM sparticles.

The first step is to find the inflaton $|Z|$ in terms of the canonical inflaton which we call $z$. We define $z$ such that $\langle z\rangle=0$. Although exact canonical parametrization may not be available (such as in our models), we only need to find the expansion of $|Z|$ up to quadratic order in $z$. Let us denote dimensionless scalars as $|\kappa Z|\equiv\rho$ and $\kappa z\equiv\tilde{z}$, and expand $\rho$ around the minimum,
\begin{equation}
    \rho=\langle\rho\rangle+\langle\rho'\rangle\tilde{z}+\tfrac{1}{2}\langle\rho''\rangle\tilde{z}^2+\ldots~,\label{rho_z}
\end{equation}
where the prime stands for $d/d\tilde{z}$. Derivatives of $\rho$ can be found by requiring canonical normalization of the inflaton,
\begin{equation}
    e^{-1}{\cal L}\supset -\kappa^{-2}G(\rho)\partial_m\rho\partial^m\rho=-\tfrac{1}{2}\kappa^{-2}\partial_m\tilde z\partial^m\tilde z~,~~~G(\rho)\equiv J_{Z\bar Z}(\rho)~.
\end{equation}
This yields
\begin{equation}
    \rho'=\frac{1}{\sqrt{2G}}~,~~~\rho''=-\frac{G'}{\sqrt{(2G)^3}}=-\frac{G_\rho\rho'}{\sqrt{(2G)^3}}=-\frac{G_\rho}{4G^2}~.\label{rho_prime_z}
\end{equation}
Then, any function $f(\rho)$ can be expanded around the minimum ($\rho=\langle \rho\rangle$ which corresponds to $\tilde z=0$) in terms of the canonical $\tilde{z}$ as follows (up to quadratic terms),
\begin{align}
\begin{split}
    f(\rho) &=\langle f\rangle+\langle f_\rho\rangle(\rho-\langle\rho\rangle)+\tfrac{1}{2}\langle f_{\rho\rho}\rangle(\rho-\langle\rho\rangle)^2\\
    &=\langle f\rangle+\frac{\langle f_\rho\rangle}{\sqrt{2\langle G\rangle}}\tilde{z}-\frac{\langle f_\rho G_\rho\rangle-2\langle Gf_{\rho\rho}\rangle}{8\langle G\rangle^2}\tilde{z}^2~,\label{f_z_expansion}
\end{split}
\end{align}
where we used \eqref{rho_z} and \eqref{rho_prime_z}.

Using these results, we can derive the interactions of the inflaton with MSSM sparticles.
\paragraph{Trilinear scalar interactions}
The trilinear scalar interactions involving the inflaton are universal, and given by
\begin{align}
e^{-1}{\cal L}\supset -C_{\phi} z\sum|\phi|^2~,~~~C_{\phi}\equiv \frac{\kappa m_{3/2}^2}{\langle\rho\rangle^2\sqrt{2\langle G\rangle}}\left\langle h_\rho+\kappa^2J_\rho h-4\rho-2\kappa^2J_\rho\rho^2\right\rangle~,\label{inf_tri_scalar}
\end{align}
where the sum is over all MSSM scalars $\phi$. We treat $J$ as a function of $\rho$ and introduce the notation
\begin{equation}
	h\equiv J^{Z\bar Z}|1+\kappa^2J_ZZ|^2~,
\end{equation}
also as a function of $\rho$. As can be seen the function $h$ is dimensionless.

\paragraph{Quartic scalar interactions}
For quartic scalar interactions we have
\begin{equation}
	e^{-1}{\cal L}\supset -d_{\phi}\left[y_uz\bar{\tilde u}\tilde QH_u-y_dz\bar{\tilde d}\tilde QH_d-y_ez\bar{\tilde e}\tilde LH_d+{\rm h.c.}\right]+\tilde{d}_{\phi}z^2\sum|\phi|^2~.\label{inf_four_scalar}
\end{equation}
where
\begin{align}
    d_{\phi} &\equiv\frac{\kappa m_{3/2}}{\langle\rho\rangle^2\sqrt{2\langle G\rangle}}\big\langle h_\rho+\kappa^2J_\rho h\big\rangle~,\\
\begin{split}
    \tilde{d}_{\phi} &\equiv \frac{\kappa^2m_{3/2}^2}{2\langle\rho^2G\rangle}\bigg\langle 2-\frac{\rho G_\rho}{G}+\frac{h_\rho G_\rho}{4G}-\frac{h_{\rho\rho}}{2}-\kappa^2J_\rho h_\rho\\
    &\qquad +4\kappa^2J_\rho\rho+\frac{\kappa^2}{2}\left(\frac{J_\rho G_\rho}{2G}-J_{\rho\rho}-\kappa^2J_{\rho}^2\right)(h-2\rho^2)\bigg\rangle~.
\end{split}
\end{align}
\paragraph{Inflaton-gaugino interactions}
The trilinear inflaton-gaugino terms are given by
\begin{equation}
	e^{-1}{\cal L}\supset \tfrac{1}{2}C_{a}z\lambda^{a}\lambda^{a}+{\rm h.c.}~,
\end{equation}
where
\begin{gather}
    C_{a}\equiv \frac{\kappa m_{3/2}\beta_a}{2\langle\rho\rangle^3\sqrt{2\langle G\rangle}}\left\langle\tfrac{1}{2}\kappa^2\rho J_\rho\hat h+\rho\hat h_\rho-\hat h\right\rangle~,~~~\hat h\equiv J^{Z\bar Z}(1+\kappa^2J_ZZ)~.
\end{gather}
There is also the following inflaton-higgsino interaction,
\begin{equation}
    e^{-1}{\cal L}=-\left\langle\frac{2+\rho \kappa^2J_\rho}{2\rho\sqrt{2G}}\right\rangle\kappa\mu z\tilde H_u\tilde H_d~,
\end{equation}
which is suppressed by $\kappa\mu$.

For model I with parameter set \eqref{model_I_parameters} the inflaton-MSSM couplings $C_\phi$, $C_{a}$, $d_\phi$, and $\tilde d_\phi$ have the values shown in Table \ref{tab_interactions_I}.

\begin{table}[hbt!]
\centering
\begin{tabular}{r r r r r r}
\toprule
$C_\phi/{\rm GeV}$ & $d_\phi$ & $\tilde d_\phi$ & $C_1$ & $C_2$ & $C_3$ \\
\hline
$8.6\times 10^{4}$ & $1.39\times 10^{-6}$ & $-1.81\times 10^{-13}$ & $2.4\times 10^{-10}$ & $1.52\times 10^{-10}$ & $1.35\times 10^{-10}$ \\\bottomrule
\hline
\end{tabular}
\captionsetup{width=1\linewidth}
\caption{Inflaton-MSSM couplings for model I with parameter set \eqref{model_I_parameters}.}
\label{tab_interactions_I}
\end{table}
\paragraph{Inflaton-inflatino interactions}
Since in our models SUSY is broken by both $F$- and $D$-terms, inflatino and gaugino mix with each other. In particular, the goldstino $\eta$ is the following combination of fermions
\begin{equation}
    \kappa\eta=\left(\kappa^2K_Z+\frac{W_Z}{W}\right)\zeta-\frac{i}{\sqrt{2}}e^{-\kappa^2K/2}\frac{{\cal D}_R}{W}\lambda^R~.\label{goldstino_1}
\end{equation}
The physical fermion can be identified after choosing the unitary gauge $\eta=0$. We can ignore the MSSM fields and use ${\cal D}_R=g(\kappa^{-2}+J_ZZ)$ and $W=\kappa^{-2}aZ$. Then, upon setting $\eta=0$ in Eq. \eqref{goldstino_1}, we have
\begin{equation}
    \lambda^R=-\frac{\sqrt{2}ia}{g}e^{\kappa^2J/2}\zeta~,\label{ortho_fermion}
\end{equation}
which can be used to express the physical fermion orthogonal to the goldstino as $\zeta$ (or $\lambda^R$).

The kinetic terms for $\zeta$ and $\lambda^R$ are given by
\begin{equation}
    e^{-1}{\cal L}\supset -iJ_{Z\bar Z}\zeta\sigma^mD_m\bar\zeta-i\lambda^R\sigma^mD_m\bar\lambda^R~=-i\left(J_{Z\bar Z}+\frac{2a^2}{g^2}e^{\kappa^2J}\right)\zeta\sigma^mD_m\bar\zeta+\ldots~,
\end{equation}
where we used \eqref{ortho_fermion} and $f_R\approx 1$. Thus, for canonical normalization around the minimum we require
\begin{equation}
    \zeta\rightarrow\left\langle J_{Z\bar Z}+\frac{2a^2}{g^2}e^{\kappa^2J}\right\rangle^{-\frac{1}{2}}\zeta~.\label{canonical_inflatino}
\end{equation}

The mass term of the physical fermion, as well as its trilinear coupling to the canonical inflaton $z$, can be derived from the following expression \cite{Wess:1992cp},
\begin{align}
\begin{split}
    e^{-1}{\cal L} &\supset \sqrt{2}\left(J_{Z\bar Z}\overbar X_R^Z-\tfrac{i}{4}\partial_Zf_R{\cal D}_R\right)\zeta\lambda^R\\
    &+\tfrac{1}{2}e^{\kappa^2J/2}\Big[(W_{ZZ}+\kappa^2J_{ZZ}W+2\kappa^2J_ZD_ZW-\kappa^4J_Z^2W-\Gamma^Z_{ZZ}D_ZW)\zeta^2\\
    &\qquad\qquad\qquad -\tfrac{1}{2}J^{Z\bar Z}D_{\bar Z}\overbar W\partial_Zf_R(\lambda^R)^2\Big]+{\rm h.c.}~,\label{zeta_lambda_R_int}
\end{split}
\end{align}
where the terms with $\partial_Zf_R$ can be ignored since they are suppressed by $g^2$ (compared to other terms). $\Gamma^Z_{ZZ}=J^{Z\bar Z}\partial_Z J_{Z\bar Z}$ is the Christoffel symbol for the K\"ahler metric $J_{Z\bar Z}$. To derive the mass term and relevant coupling, we use Eq.~\eqref{ortho_fermion}, the explicit forms of $X^Z_R, {\cal D}_R, W$, and the canonical normalization \eqref{canonical_inflatino}. Then Eq.~\eqref{zeta_lambda_R_int} becomes
\begin{gather}
    e^{-1}{\cal L}\supset -\frac{a}{2\kappa}j\zeta^2+{\rm h.c.}~,\\
    j\equiv \kappa e^{\kappa^2J/2}\left(J_{Z\bar Z}+\frac{2a^2}{g^2}e^{\kappa^2J}\right)^{-1}\left[2J_Z-\kappa^{-2}\Gamma^Z_{ZZ}+(J_{ZZ}+\kappa^2J_Z^2-J_Z\Gamma^Z_{ZZ})Z-4J_{Z\bar Z}\overbar Z\right]~,
\end{gather}
where $j$ is dimensionless.

Finally, expanding $j$ in terms of the canonical inflaton $z$ by using Eq.~\eqref{f_z_expansion} we find
\begin{equation}
    e^{-1}{\cal L}\supset \frac{1}{2}\left(-\frac{a\langle j\rangle}{\kappa}+C_\zeta z\right)\zeta^2+{\rm h.c.}~,~~~C_\zeta\equiv -\frac{a\langle j_\rho\rangle}{\sqrt{2\langle J_{Z\bar Z}\rangle}}~,\label{zeta_bilinear_term}
\end{equation}
where $\rho$ is the dimensionless non-canonical inflaton $|\kappa Z|$, as before.
The first term in \eqref{zeta_bilinear_term} is the mass term of the inflatino $m_\zeta=a\langle j\rangle/\kappa$ which is $6.15\times 10^{11}$ GeV for the parameter set \eqref{model_I_parameters}, as we mentioned at the beginning of this section. The second term of \eqref{zeta_bilinear_term} is the trilinear inflaton-inflatino coupling. For the parameter choice \eqref{model_I_parameters} its value is $C_\zeta=9.67\times 10^{-8}$.

\subsection{Reheating}

In model I the inflaton $z$ can perturbatively decay into the MSSM scalars, gaugini, and inflatino since their masses are smaller than $m_z/2$. 

The decay channels into MSSM scalars are given by Eqs. \eqref{inf_tri_scalar} and \eqref{inf_four_scalar}. In the quartic interactions the dominant contribution comes from the stop part due to its Yukawa coupling $y_t={\cal O}(1)$ (other Yukawa couplings are much smaller). Therefore the relevant terms for reheating are
\begin{equation}
    e^{-1}{\cal L}\supset -C_\phi z\sum|\phi|^2-d_\phi z(y_t\bar{\tilde t}\tilde Q_3H_u+{\rm h.c.})~,\label{z_tri_four}
\end{equation}
where $\tilde Q_3$ is the third-generation quark doublet.

Three-point ($-\sigma z\varphi^2$) and four-point ($-\lambda z\varphi^3$) decay rates for scalar particles (both $z$ and $\varphi$ real) are given by \cite{ParticleDataGroup:2020ssz}
\begin{equation}
    \Gamma_{z\rightarrow \varphi\varphi}=\frac{\sigma^2}{8\pi m_z}~,~~~\Gamma_{z\rightarrow\varphi\varphi\varphi}=\frac{\lambda^2m_z}{3!64(2\pi)^3}~,
\end{equation}
ignoring the masses of the final-state particles (which can be justified in our case since the MSSM scalars are lighter than the inflaton roughly by a factor of ten).

After including both real degrees of freedom canonically normalized as $\phi=(\phi_1+i\phi_2)/\sqrt{2}$ ($z$ is already real canonical), the total decay rate into two MSSM scalars is
\begin{equation}
    \Gamma_{z\rightarrow\phi\phi}^{\rm tot}=\frac{C^2_\phi}{16\pi m_z}\cdot 49=5.77\times 10^{-3}~{\rm GeV}~,
\end{equation}
where $49$ is the number of species.

The decay rate into three scalars comes from the second term of \eqref{z_tri_four}, which expands into three color and two weak components. In general, the product of three complex scalars plus its Hermitian conjugate reads, in terms of their real and imaginary parts, for example $A=(A_1+iA_2)/\sqrt{2}$,
\begin{equation}
    ABC+{\rm c.c.}=\frac{1}{\sqrt{2}}(A_1B_1C_1-A_1B_2C_2-A_2B_1C_2-A_2B_2C_1)~,
\end{equation}
i.e. each individual quartic interaction of \eqref{z_tri_four} contains four terms. Taking all of this into account, the total decay rate of $z$ into three MSSM scalars (six relevant species) is given by
\begin{equation}
    \Gamma_{z\rightarrow\phi\phi\phi}^{\rm tot}=\frac{d^2_\phi y_t^2 m_z}{192(2\pi)^3}\cdot 6=3.04\times 10^{-4}~{\rm GeV}~,
\end{equation}
where we take $y_t=1$ for simplicity.

For the coupling with two fermions of the form $\tfrac{1}{2}C_\lambda z\lambda^2+{\rm h.c.}$, the decay rate is given by
\begin{equation}
    \Gamma_{z\rightarrow \lambda\lambda}=\frac{C^2_\lambda m_z}{8\pi}~.
\end{equation}
The values of the individual decay rates into gaugini and inflatino are shown in Table \ref{tab_model_I_decay} where it can be seen that the decay into gaugini is negligible in comparison to the decay into inflatino. The total decay rate into fermions is then
\begin{equation}
    \Gamma^{\rm tot}_{z\rightarrow\lambda\lambda}\approx\Gamma_{z\rightarrow \zeta\zeta}=4.65\times 10^{-4}~{\rm GeV}~.
\end{equation}

\begin{table}[hbt!]
\centering
\begin{tabular}{r r r r}
\toprule
$\Gamma_{z\rightarrow \lambda_1\lambda_1}$ & $3\Gamma_{z\rightarrow \lambda_2\lambda_2}$ & $8\Gamma_{z\rightarrow \lambda_3\lambda_3}$ & $\Gamma_{z\rightarrow \zeta\zeta}$ \\
\hline
$2.86\times 10^{-9}$ & $3.45\times 10^{-9}$ & $7.25\times 10^{-9}$ & $4.65\times 10^{-4}$ \\\bottomrule
\hline
\end{tabular}
\captionsetup{width=1\linewidth}
\caption{Inflaton decay rates (in GeV) into gaugini and inflatino for model I with parameter set \eqref{model_I_parameters}.}
\label{tab_model_I_decay}
\end{table}

From the total decay rate
\begin{equation}
    \Gamma_{\rm tot}=\Gamma^{\rm tot}_{z\rightarrow\phi\phi}+\Gamma^{\rm tot}_{z\rightarrow\phi\phi\phi}+\Gamma^{\rm tot}_{z\rightarrow\lambda\lambda}=6.54\times 10^{-3}~{\rm GeV}~,
\end{equation}
we can estimate the reheating temperature as,
\begin{equation}
    T_{\rm reh}\simeq \sqrt{M_P \Gamma_{\rm tot}}=1.26\times 10^{8}~{\rm GeV}~.
\end{equation}

If we take other parameter values ($\alpha,\beta,g$) from the allowed region in Figure \ref{Fig_par_space}, the mass hierarchy and the results for reheating may change, for example for
\begin{equation}
    \alpha=0.007~,~~~\beta=0.0499~,~~~g/a=0.1~,\label{model_I_par_other}
\end{equation}
the VEV of the inflaton is pushed to $\langle\kappa Z\rangle=1.2$, and the observables are $n_s=0.9631$, $r=7.86\times 10^{-5}$ ($a$ is fixed as $1.56\times 10^{-6}$). The corresponding mass spectrum is shown in Table \ref{tab_soft_I_other}. In addition, the $U(1)_R$ vector mass is $9.23\times 10^{11}$ GeV. It can be seen that except for the inflatino and the $U(1)_R$ vector, all the masses are larger than in our previous parameter choice (see Table \ref{tab_soft_I}) but more importantly, the decay of the inflaton into MSSM scalars is kinematically forbidden, $m_z<2m_0$, and the dominant contribution to the reheating temperature comes from MSSM gaugino decays. We found that the inflaton-gaugino couplings are larger in this case, which ultimately leads to a reheating temperature of the same order as before, at $T_{\rm reh}\sim 10^{8}$ GeV.

\begin{table}[hbt!]
\centering
\begin{tabular}{r r r r r r r r}
\toprule
 $m_z$ & $m_\zeta$ & $m_{3/2}$ & $m_0$ & $m_1$ & $m_2$ & $m_3$ \\
\hline
$1.48\times 10^{13}$ & $4.07\times 10^{10}$ & $1.05\times 10^{13}$ & $1.04\times 10^{13}$ & $1.84\times 10^{11}$ & $1.16\times 10^{11}$ & $1.04\times 10^{11}$ \\\bottomrule
\hline
\end{tabular}
\captionsetup{width=1\linewidth}
\caption{Mass spectrum of model I with parameter set \eqref{model_I_par_other} (in GeV).}
\label{tab_soft_I_other}
\end{table}

\section{Model II: hyperbolic K\"ahler space}

The second inflationary model (with gauged $R$-symmetry) that we would like to consider is given by the K\"ahler potential \cite{Aldabergenov:2020bsl}
\begin{equation}
    J=-\kappa^{-2}\gamma\log(1-|\kappa Z|^2-\delta|\kappa Z|^4)~,
\end{equation}
with two subclasses defined by $\gamma=2$ and $\gamma=3$, while $\delta$ is some real parameter which must satisfy $|\delta|\ll 1$ in order to describe slow-roll inflation (the parameters $\gamma$ and $\delta$ are called $\alpha$ and $\beta$ in \cite{Aldabergenov:2020bsl}, while our $a$ is denoted by $\mu$ in that work). The superpotential for the inflaton is the same as before, $W=a\kappa^{-2}Z$, and the scalar potential of this model is given by
\begin{align}
    \kappa^{-4}V=a^2(1-|\kappa Z|^2-\delta|\kappa Z|^4)^{-\gamma}&\left\{\frac{\left(1+(\gamma-1)|\kappa Z|^2+\delta(2\gamma-1)|\kappa Z|^4\right)^2}{\gamma(1+4\delta|\kappa Z|^2-\delta|\kappa Z|^4)}-3|\kappa Z|^2\right\}\nonumber\\
    &+\frac{g^2}{2}\left\{\frac{1+(\gamma-1)|\kappa Z|^2+\delta(2\gamma-1)|\kappa Z|^4}{1-|\kappa Z|^2-\delta|\kappa Z|^4}\right\}^2~.\label{model_II_potential}
\end{align}
As was described in \cite{Aldabergenov:2020bsl}, when $\delta=0$ we obtain no-scale de Sitter potential for $\gamma=2$ and $a^2=2g^2$, and for $\gamma=3$ and $a^2=9g^2/2$ (in both of these cases, the potential becomes constant and proportional to $g^2$). For realistic inflation we need a small deviation from these relations between $a$ and $g$ as well as a small non-zero $|\delta|$.

First let us fix $\gamma$. We find that when $\gamma=2$, it is difficult to obtain positive squared masses for the MSSM scalars without extra scalars in the hidden sector. In particular we have,
\begin{equation}
	m^2_0=m^2_{3/2}-\tfrac{\kappa^2}{2}\langle{\cal D}_R\rangle^2<0~,
\end{equation}
for the parameter sets described in \cite{Aldabergenov:2020bsl}. On the other hand when $\gamma=3$, the squared mass $m^2_0$ is positive even without extra scalars. To keep the field content minimal, we choose the $\gamma=3$ case.

Next, as mentioned above, for $\gamma=3$ we need a small deviation from $\delta=0$ and from $a^2=9g^2/2$ in order to describe inflation. In particular, it is convenient to introduce a new parameter $\lambda$ defined as 
\begin{equation}
    a^2=\tfrac{9}{2}g^2(1+\lambda)~.\label{ag_rel}
\end{equation}
Then we require $|\delta|\ll 1$ and $|\lambda|\ll 1$. Slow-roll inflation and stable Minkowski vacuum are possible along a specific trajectory in $\delta-\lambda$ space (with $a$ and $g$ fixed by the scale of inflationary perturbations). This trajectory can be seen in Figure 6 of Ref. \cite{Aldabergenov:2020bsl}.

Here as a specific example we take the following parameter values,~\footnote{The parameter $\delta$ has no lower bound, except that it cannot vanish, but we find that smaller $\delta$ leads to larger soft parameters (as well as larger $n_s$). In particular, the dimensionful soft parameters can exceed Planck mass if $\delta$ is too small, e.g. for $\delta=10^{-3}$ the gravitino mass becomes of order Planck mass. For this reason we choose a relatively large value of $\delta$.}
\begin{equation}
\delta=0.03~,~~~\lambda=-3.991\times 10^{-2}~,~~~g=4.79\times 10^{-6}~,\label{model_II_parameters}
\end{equation}
where the limit $\delta=\lambda=0$ leads to the no-scale case $V=2g^2/\kappa^4$. 

\begin{figure}
\centering
  \centering
  \includegraphics[width=.45\linewidth]{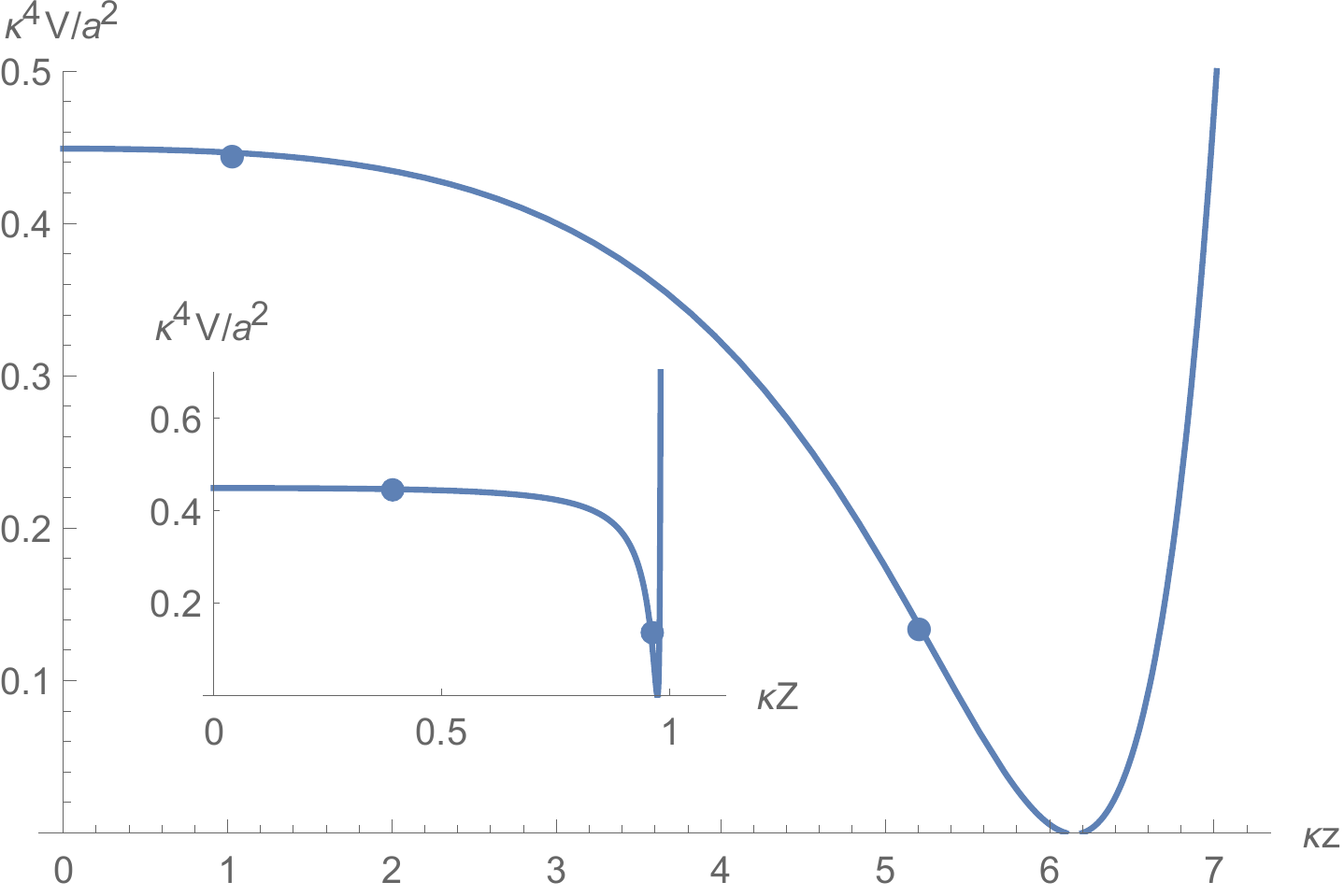}
\captionsetup{width=1\linewidth}
\caption{Scalar potential of model II for the parameter set \eqref{model_II_parameters}. Both canonical (main plot) and non-canonical (subplot) parametrizations are shown. The markers represent start and end of $60$ e-folds of inflation.}\label{Fig_V_2}
\end{figure}

The scalar potential for the parameters \eqref{model_II_parameters} is displayed in Figure \ref{Fig_V_2} where we show both the non-canonical original scalar $Z$ and the canonical one $z$ (which is found numerically). In contrast to the previous model, here we always have positive $m_0^2$ in the allowed parameter space.

In this model the inflationary parameters are given by
\begin{equation}
	H=9.38\times 10^{12}~{\rm GeV}~,~~~n_s=0.9576~,~~~r=1.5\times 10^{-3}~,
\end{equation}
with the inflaton mass $m_z=1.93\times 10^{13}$ GeV, the vector mass $m_A=1.1\times 10^{15}$ GeV, and the inflatino mass $m_\zeta=1.1\times 10^{14}$ GeV.

Since we already derived MSSM soft parameters and inflaton-MSSM interactions for general $J$, here we just show the explicit results taking the parameter set \eqref{model_II_parameters}. The mass spectrum is summarized in Table \ref{tab_soft_II} and the values of the inflaton-MSSM coupling constants in Table \ref{tab_interactions_II}.

\begin{table}[hbt!]
\centering
\begin{tabular}{r r r r r r r}
\toprule
$m_{z}$ & $m_{\zeta}$ & $m_{3/2}$ & $m_0$ & $m_1$ & $m_2$ & $m_3$ \\
\hline
$1.93\times 10^{13}$ & $1.1\times 10^{14}$ & $5.55\times 10^{15}$ & $5.47\times 10^{15}$ & $2.49\times 10^{12}$ & $1.57\times 10^{12}$ & $1.41\times 10^{12}$ \\\bottomrule
\hline
\end{tabular}
\captionsetup{width=1\linewidth}
\caption{MSSM soft parameters (and the gravitino mass) derived from model II with parameter set \eqref{model_II_parameters}. All the parameters are in GeV. The Standard Model gauge coupling values are taken at $10^8$ GeV.}
\label{tab_soft_II}
\end{table}

\begin{table}[hbt!]
\centering
\begin{tabular}{r r r r r r}
\toprule
$C_\phi/{\rm GeV}$ & $d_\phi$ & $\tilde d_\phi$ & $C_1$ & $C_2$ & $C_3$ \\
\hline
$3.02\times 10^{13}$ & $1.66\times 10^{-2}$ & $-1.52\times 10^{-5}$ & $2.07\times 10^{-7}$ & $1.31\times 10^{-7}$ & $1.17\times 10^{-7}$ \\\bottomrule
\hline
\end{tabular}
\captionsetup{width=1\linewidth}
\caption{Inflaton-MSSM couplings for model II with parameter set \eqref{model_II_parameters}.}
\label{tab_interactions_II}
\end{table}

In our model II, MSSM scalars and inflatino are heavier than the inflaton, which means that the inflaton can only (perturbatively) decay into the gaugini $\lambda_{1,2,3}$. The individual decay rates are
\begin{align}
    \Gamma_{z\rightarrow \lambda_1\lambda_1} &=3.29\times 10^{-2}\,{\rm GeV}~,\nonumber\\
    3\Gamma_{z\rightarrow \lambda_2\lambda_2} &=3.95\times 10^{-2}\,{\rm GeV}~,\\
    8\Gamma_{z\rightarrow \lambda_3\lambda_3} &=8.41\times 10^{-2}\,{\rm GeV}~,\nonumber
\end{align}
and the total decay rate is $\Gamma_{\rm tot}=0.16$ GeV. The reheating temperature is estimated as $T_{\rm reh}\simeq 6\times 10^{8}$ GeV.

As can be seen from Table \ref{tab_soft_II}, the gravitino in this case is much heavier than the inflaton, which prohibits the perturbative decay of the inflaton into two gravitini. In fact, this remains true for the whole parameter range that is suitable for realistic inflation, which can be seen as follows. In \cite{Aldabergenov:2020bsl} it was shown that the scalar potential (for the canonical inflaton $z$) of model II in the aforementioned parameter range, can be approximated by the Starobinsky potential,
\begin{equation}
    V(z)\simeq \frac{2g^2}{\kappa^4}(1-e^{-\sqrt{\frac{2}{3}}\kappa z})^2~,
\end{equation}
which means that the inflaton mass is $m_z\approx \sqrt{8/3}\kappa^{-1}g$. Here $g$ is our gauge coupling as before, which is related to $a$ by \eqref{ag_rel}, so that when $|\lambda|\ll 1$ we have $g\approx\sqrt{2}a/3$. On the other hand, the gravitino mass is given by $m_{3/2}=a\langle e^{\kappa^2J/2}|Z|\rangle$. In the relevant parameter region the VEV of $Z$ is $|\langle Z\rangle|\approx \kappa^{-1}$ (for example for the parameter choice \eqref{model_II_parameters} we have $|\langle Z\rangle|\approx 0.97\kappa^{-1}$), which leads to large VEV of $J$,
\begin{equation}
    \kappa^2\langle J\rangle=-3\log(1-|\langle\kappa Z\rangle|^2-\delta|\langle\kappa Z\rangle|^4)\gg 1~,
\end{equation}
where, again, $|\delta|\ll 1$. This implies that $m_z\ll m_{3/2}$. More specifically, we find that in the relevant parameter space $m_z$ is smaller than $m_{3/2}$ by at least two orders of magnitude.

\section{Modifications of K\"ahler potential}

Unlike superpotential, K\"ahler potential in $N=1$ theories is not protected from renormalization effects if they are allowed by symmetries of the model. In our case the local $U(1)_R$ symmetry allows for interaction terms between matter fields $\phi$ and the inflaton $Z$, as in the K\"ahler potential
\begin{equation}
    K=\phi\bar\phi+J(Z\overbar Z)+\phi\bar\phi\Delta(Z\overbar Z)~,\label{K_Delta}
\end{equation}
where $\Delta(Z\overbar Z)$ is an arbitrary function of $Z\overbar Z$. We take one matter field for simplicity, but for each one we can introduce a different invariant function $\Delta_i(Z\overbar Z)$.

Let us now examine the stability of the matter scalar potential in our models I and II, w.r.t. the modifications of the K\"ahler potential of the form \eqref{K_Delta}. In particular, we want to make sure that the soft scalar mass squared, $m_0^2$, remains positive at the minimum. Previously we found that $m_0^2$ can be expressed as (for general $J$-function)
\begin{equation}
    m_0^2=m^2_{3/2}-\tfrac{\kappa^2}{2}\langle {\cal D}_R\rangle^2~,
\end{equation}
as found in Subsection \ref{subsec_m0}. Once we take into account the $\Delta$-term in the K\"ahler potential \eqref{K_Delta}, this expression is modified as
\begin{equation}
    m_0^2=m^2_{3/2}-\tfrac{\kappa^2}{2}\langle 1+\Delta\rangle\langle {\cal D}_R\rangle^2~.
\end{equation}
Therefore if $\langle\Delta\rangle\leq 0$,
it cannot destabilize the $\langle\phi\rangle=0$ vacuum (assuming we have $m_0^2>0$ in the absence of $\Delta$). On the other hand, positive $\langle\Delta\rangle$ may introduce a tachyonic instability in $\phi$-direction, if it is too large. This leads to the upper bound on the value of positive $\langle\Delta\rangle$, which we call $\langle\Delta\rangle_{\rm max}$, depending on a particular model and its parameters. For example in model I with parameter set \eqref{model_I_parameters} the upper bound is $\langle\Delta\rangle_{\rm max}\approx 0.147$. In model II with parameter choice \eqref{model_II_parameters} we find $\langle\Delta\rangle_{\rm max}\approx 33$.

Obviously, if MSSM scalars have distinct $\Delta$-terms, this will lead to splitting between their masses,
\begin{equation}
    m_{0,i}^2=m^2_{3/2}-\tfrac{\kappa^2}{2}\langle 1+\Delta_i\rangle\langle {\cal D}_R\rangle^2~.
\end{equation}

\section{Dark matter candidates}

In both of the discussed models we have the mass hierarchy
\begin{equation}
    \{m_z,m_{3/2},m_0\}>\{m_1,m_2,m_3\}~,
\end{equation}
while the inflatino mass can be on either side depending on the parameter choice. On the other hand we assume that the parameter $\mu$ is much smaller than gaugino masses (for example from $100$ GeV to TeV range) to avoid extreme fine-tuning of the Higgs boson mass. This leads to the so-called split Higgsino scenario \cite{Co:2021ion} (see also \cite{Hall:2011jd}) where the two lightest neutralino states are Higgsino-like (the parameter $\mu$ corresponds exactly to the Higgsino Dirac mass term $-\mu\tilde H_u\tilde H_d$). This makes the Higgsino-like LSP a potential candidate for the thermal cold dark matter if its mass lies within the TeV range. However, if $\mu$ is around the TeV scale, the huge difference between gaugino masses and $\mu$ suppresses the mass difference $\Delta m$ between LSP and next-to-LSP. This, in turn, may lead to inelastic scattering with nuclei which is constrained by direct detection experiments \cite{Tucker-Smith:2001myb,Tucker-Smith:2004mxa}. In Ref. \cite{Nagata:2014wma} the authors showed that the lower bound on the mass difference $\Delta m\gtrsim 200$ keV translates into the upper bound on the gaugino masses $\{m_1,m_2\}\lesssim {\cal O}(10^{7})$ GeV, if we assume that Higgsino-like LSP is the dominant part of dark matter. Therefore, as long as one persists in TeV scale $\mu$, Higgsino-like LSP as the thermal dark matter is excluded in our models, because gaugino masses are much larger than the aforementioned upper bounds (the exception from this would be non-neutralino LSP scenarios, such as a hidden sector LSP or gravitino LSP, although the latter is not applicable to our models since our gravitino is generally too heavy). The overproduction of Higgsino dark matter also excludes a large window of its mass $\mu$, from TeV scale all the way up to the reheating temperature (assuming standard thermal history), which in our case is at least $10^8$ GeV. 

If we give up the fine-tuning arguments and make $\mu$ as large as the SUSY breaking scale, LSP can become a candidate for superheavy dark matter \cite{Chung:1998rq,Berezinsky:2008bg} with mass $m_{\rm LSP}\sim 10^{10}$ GeV.~\footnote{In our models, if $\mu$ is close to the inflationary scale or SUSY breaking scale (both at around $10^{12}$ GeV), depending on the particular parameter choice, including the choice of $\mu$, LSP can be bino-, wino-, Higgsino-like neutralino, or inflatino (more precisely combination of inflatino and $U(1)_R$-gaugino).} In \cite{Chung:1998rq} it was shown that such heavy particles can be produced in sufficient amounts to describe dark matter, even if their mass exceeds the reheating temperature $T_{\rm reh}$, since the maximum temperature during reheating can exceed $T_{\rm reh}$ by several orders of magnitude. Analytical approximation of the dark matter (denoted $X$) abundance in this case is given by \cite{Chung:1998rq}
\begin{equation}
    \Omega_Xh^2\simeq \alpha_X\left(\frac{g_*}{200}\right)^{-\frac{3}{2}}\left(\frac{2000 T_{\rm reh}}{m_X}\right)^7~,
\end{equation}
where $\alpha_X\equiv m_X^2\langle\sigma|\upsilon|\rangle$ with $\langle\sigma|\upsilon|\rangle$ the thermally averaged annihilation cross-section of $X$, $m_X$ is its mass, and $g_*$ is the number of relativistic degrees of freedom. Taking $g_*=100$, and requiring $\Omega_X^2h^2\simeq 0.3$ (all dark matter is in $X$ particles) we obtain a required value of the ratio $T_{\rm reh}/m_X$,
\begin{equation}
    \frac{T_{\rm reh}}{m_X}\simeq 5\times 10^{-4}\left(\frac{0.1}{\alpha_X}\right)^{\frac{1}{7}}~,
\end{equation}
to produce the right amount of $X$ dark matter. As can be seen, the result is not very sensitive to the annihilation parameter $\alpha_X$. For example as long as $\alpha_X\gtrsim 10^{-6}$, the ratio is roughly $T_{\rm reh}/m_X\sim 10^{-3}$. Let us assume that superheavy neutralino in our model I is one of the gaugini with the mass $m_X\sim 10^{10}$ GeV (see Table \ref{tab_soft_I}). Since the reheating temperature in this case is $10^8$ GeV, this does not lead to the correct dark matter abundance (overproduction). It is however possible to obtain the correct ratio $T_{\rm reh}/m_X$ if we choose different parameters. For example the parameter choice \eqref{model_I_par_other} leads to a more suitable gaugino masses $\sim 10^{11}$ GeV, as can be seen from Table \ref{tab_soft_I_other}, while the reheating temperature remains the same. We conclude that the parameter space in our models is flexible enough to accommodate superheavy LSP dark matter.

\section{Discussion}

In this paper we considered a class of single-field inflationary models defined by the gauging of $U(1)_R$ symmetry, and spontaneous breaking of supersymmetry and the $R$-symmetry after inflation where the goldstino is associated with a combination of inflatino and $U(1)_R$ gaugino. We focused on two subclasses of these models, which we call model I and model II, characterized by the geometry of the K\"ahler space. Model I has the canonical K\"ahler potential with higher-order corrections,
\begin{equation}
    K=|Z|^2+\alpha\kappa^2|Z|^4+\beta\kappa^4|Z|^6~,\label{dis_KI}
\end{equation}
while model II has hyperbolic geometry $SU(1,1)/U(1)$, also including corrections,
\begin{equation}
    K=-3\log(1-|\kappa Z|^2-\delta |\kappa Z|^4)~,\label{dis_KII}
\end{equation}
where $|Z|$ plays the role of the inflaton, while the phase of $Z$ combines with the $U(1)_R$ gauge field to form a massive vector, with the mass close to the inflationary scale in model I, and far exceeding it in model II. The superpotential is fixed by the $R$-symmetry as $W=a\kappa^{-2}Z$. These models can describe slow-roll inflation with $n_s$ and $r$ within $2\sigma$ CMB constraints ($n_s\gtrsim 0.954$, $r\lesssim 0.064$). For example for our reference parameter values \eqref{model_I_parameters} and \eqref{model_II_parameters}, we have $n_s=0.9543$ and $r=1.72\times 10^{-6}$ for model I, and $n_s=0.9576$ and $r=1.5\times 10^{-3}$ for model II.

In both models the non-canonical inflaton $|Z|$ takes sub-Planckian values, starting around $|Z|=0$ and settling at $|\langle Z\rangle|\simlt M_P$ after inflation. This means that for small parameters $\alpha,\beta,\delta$ in \eqref{dis_KI} and \eqref{dis_KII}, we can treat the correction terms in the K\"ahler potentials as perturbations. On the other hand, the canonically normalized inflaton $z$ travels sub-Planckian distances in model I (Figure \ref{Fig_V_1}), but super-Planckian distances in model II (Figure \ref{Fig_V_2}).

Having established the inflationary stage, we then coupled these models to MSSM according to Eqs. \eqref{K_master} and \eqref{W_master}, and derived the resulting soft parameters and mass spectrum (shown in Tables \ref{tab_soft_I} and \ref{tab_soft_II}). In this minimalistic approach, the scale of inflationary scalar perturbations fixes the scale of SUSY breaking, and therefore the scale of MSSM soft parameters. In particular, the universal soft scalar mass $m_0$, as well as the scalar couplings $A_0$ and $B_0$, are fixed at the tree level, while the gaugino masses are obtained after implementing Green--Schwarz mechanism for cancelling one-loop anomalies due to non-vanishing fermion $R$-charges. Our examples demonstrate that MSSM gaugino masses tend to be smaller than the MSSM scalar masses (for both models), but the mass hierarchy of the hidden sector fields (inflaton, inflatino, and the $U(1)_R$ vector) is model-dependent.

We derived the MSSM-inflaton couplings, and estimated the reheating temperature, generally $T_{\rm reh}\sim 10^{8}$ GeV, from perturbative decay channels of the inflaton: in model I it can generally decay into all the MSSM sparticles, while in model II only gaugino channels are available kinematically. Our results also show that in the interesting parameter range of both models, the inflaton mass is smaller than two times the gravitino mass, which prohibits perturbative decay of the former into two gravitini. The full picture of reheating, however, requires further investigation after taking into account non-perturbative effects such as Bose condensation and possible resonant production of fermions.

We would like to point out that the $R$-charge assignment that we used in our examples (see the superpotential of Eq. \eqref{W_master}), where the MSSM scalars are neutral, is not the only possibility. Alternatively, we can assign $R$-charge of $1/2$ to squarks and sleptons (in the convention where the superpotential has unit $R$-charge), while the Higgs scalars are neutral. In this case the quarks and leptons are neutral under $U(1)_R$. Then, if the inflaton $R$-charge is one (same as before), we have the following superpotential,
\begin{equation}
    W=(\kappa^{-3}a+\hat\mu H_uH_d)\kappa Z+\hat{y}_u\bar{u}Q H_u-\hat{y}_d\bar{d}Q H_d-\hat{y}_e\bar{e}L H_d~.
\end{equation}
Note that in contrast to our previous choice of the superpotential -- Eqs.~\eqref{W_master} and \eqref{Omega_master} -- the inflaton $Z$ does not couple to Yukawa terms here. This change of the superpotential does not significantly modify our results -- the only part affected is the MSSM gaugino masses, as they depend on the number of $R$-charged fermions which has now been reduced. The consequence of this is that the gaugino masses become smaller, but by a factor of ten at most.

Finally, we showed that our minimal models do not allow for thermal LSP dark matter, but superheavy LSP dark matter (e.g. neutralino) is possible depending on the parameter choice.

\section*{Acknowledgements}

This work was partially supported by CUniverse research promotion project of Chulalongkorn University (grant CUAASC), Thailand Science research and Innovation Fund Chulalongkorn University CU$\_$FRB65$\_$ind (2)$\_$107$\_$23$\_$37, and partially performed by I.A. as International professor of the Francqui Foundation, Belgium. We thank Daris Samart and Spyros Sypsas for useful discussions.

\section*{Appendix: Supergravity Lagrangian}
\addcontentsline{toc}{section}{\protect\numberline{}Appendix: Supergravity Lagrangian}

We use notations and conventions of Ref. \cite{Wess:1992cp}, except that we include the gauge couplings in the corresponding Killing vectors. The relevant part of $N=1$ supergravity Lagrangian reads ($M_P=1$)
\begin{align}
    \begin{split}
        e^{-1}{\cal L}_{\rm bos}&=\tfrac{1}{2}R-K_{I\bar J}D_m\Phi^ID^m\overbar\Phi^J-\tfrac{1}{4}{\rm Re}(f_{AB})F^A_{mn}F^{mnB}+\tfrac{1}{4}{\rm Im}(f_{AB})F^A_{mn}\tilde F^{mnB} \\
        &\qquad
        -e^{K}\left\{K^{I\bar{J}}D_IW D_{\bar{J}}\overbar{W}-3|W|^2\right\}-\frac{1}{2}{\rm Re}(f^{AB}){\cal D}_{A}{\cal D}_{B}~,
    \end{split}
\end{align}
for the bosonic sector, and
\begin{align}
    \begin{split}
        &e^{-1}{\cal L}_{\rm fermi}=
        \varepsilon^{k\ell mn}\bar\psi_k\bar\sigma_\ell D_m\psi_n-e^{K/2}\left(\overbar{W}\psi_m\sigma^{mn}\psi_n+W\bar\psi_m\bar\sigma^{mn}\bar\psi_n\right)\\
        &~~
        -\tfrac{i}{2}K_{I\bar J}\chi^I\sigma^mD_m\bar\chi^J-\tfrac{i}{2}{\rm Re}(f_{AB})\lambda^A\sigma^mD_m\bar\lambda^B+\sqrt{2}K_{I\bar J}\overbar X^J_A\chi^I\lambda^A\\
        &~~
        -\tfrac{i}{4}\sqrt{2}\partial_If_{AB}{\cal D}^A\chi^I\lambda^B-e^{K/2}\left(\tfrac{1}{2}D_{IJ}W\chi^I\chi^J-\tfrac{1}{4}K^{I\bar J}D_{\bar J}\overbar W\partial_If_{AB}\lambda^A\lambda^B\right)+{\rm h.c.}+\ldots~,
    \end{split}
\end{align}
for fermions (h.c. applies to the second and third lines, and $\ldots$ stands for irrelevant terms such as non-renormalizable interactions). Here
\begin{equation}
    D_{IJ}W\equiv W_{IJ}+K_{IJ}W+K_ID_JW+K_JD_IW-K_IK_JW-\Gamma^K_{IJ}D_KW~,
\end{equation}
and $D_m$ acting on the fermions are appropriate Lorentz-/K\"ahler-/gauge-covariant derivatives.

The auxiliary $F$-field is eliminated via its equation of motion,
\begin{equation}
    F^I=-e^{\kappa^2K/2}K^{I\bar J}D_{\bar J}\overbar W~,
\end{equation}
while the $D$-field is equal to Killing potential ${\cal D}_A$ (up to a minus sign) which is given by
\begin{equation}
    {\cal D}_A=iX^I_A\left(K_I+\kappa^{-2}\frac{W_I}{W}\right)~.
\end{equation}


\begin{thebibliography}{99}

\bibitem{Antoniadis:2017gjr}
I.~Antoniadis, A.~Chatrabhuti, H.~Isono and R.~Knoops,``Inflation from Supersymmetry Breaking,''
Eur. Phys. J. C \textbf{77} (2017) no.11, 724
[arXiv:1706.04133 [hep-th]].

\bibitem{Antoniadis:2019dpm}
I.~Antoniadis, A.~Chatrabhuti, H.~Isono and R.~Knoops,
``A microscopic model for inflation from supersymmetry breaking,''
Eur. Phys. J. C \textbf{79} (2019) no.7, 624
[arXiv:1905.00706 [hep-th]].

\bibitem{Antoniadis:2015mna}
I.~Antoniadis and R.~Knoops,
``MSSM soft terms from supergravity with gauged R-symmetry in de Sitter vacuum,''
Nucl. Phys. B \textbf{902} (2016), 69-94
[arXiv:1507.06924 [hep-ph]].

\bibitem{Antoniadis:2015adn}
I.~Antoniadis and R.~Knoops,
``Gauging MSSM global symmetries and SUSY breaking in de Sitter vacuum,''
Nucl. Phys. B \textbf{903} (2016), 304-324
[arXiv:1511.04283 [hep-ph]].

\bibitem{Wess:1992cp}
 J. Wess and J. Bagger, Supersymmetry and supergravity. Princeton University Press, Princeton, NJ, USA, 1992.

\bibitem{Akrami:2018odb}
\textbf{Planck} Collaboration, Y.~Akrami \textit{et al.} , ``Planck 2018 results. X. Constraints on inflation,''
Astron. Astrophys. \textbf{641} (2020), A10
[arXiv:1807.06211 [astro-ph.CO]].
 
 \bibitem{Freedman:2005up}
D.~Z.~Freedman and B.~Kors,
``Kaehler anomalies in Supergravity and flux vacua,''
JHEP \textbf{11} (2006), 067
[arXiv:hep-th/0509217]. 

\bibitem{Elvang:2006jk}
H.~Elvang, D.~Z.~Freedman and B.~Kors,
``Anomaly cancellation in supergravity with Fayet-Iliopoulos couplings,''
JHEP \textbf{11} (2006), 068
[arXiv:hep-th/0606012].
 
 \bibitem{Antoniadis:2014iea}
I.~Antoniadis, D.~M.~Ghilencea and R.~Knoops,
``Gauged R-symmetry and its anomalies in 4D N=1 supergravity and phenomenological implications,''
JHEP \textbf{02} (2015), 166
[arXiv:1412.4807 [hep-th]].

\bibitem{Kallosh:1999jj}
R.~Kallosh, L.~Kofman, A.~D.~Linde and A.~Van Proeyen,
Phys. Rev. D \textbf{61} (2000), 103503
doi:10.1103/PhysRevD.61.103503
[arXiv:hep-th/9907124 [hep-th]].

\bibitem{Addazi:2016bus}
A.~Addazi and M.~Y.~Khlopov,
Phys. Lett. B \textbf{766} (2017), 17-22
doi:10.1016/j.physletb.2016.12.044
[arXiv:1612.06417 [gr-qc]].

\bibitem{Hasegawa:2017hgd}
F.~Hasegawa, K.~Mukaida, K.~Nakayama, T.~Terada and Y.~Yamada,
Phys. Lett. B \textbf{767} (2017), 392-397
doi:10.1016/j.physletb.2017.02.030
[arXiv:1701.03106 [hep-ph]].

\bibitem{ParticleDataGroup:2020ssz}
P.~A.~Zyla \textit{et al.} [Particle Data Group],
``Review of Particle Physics,''
PTEP \textbf{2020} (2020) no.8, 083C01

\bibitem{Aldabergenov:2020bsl}
Y.~Aldabergenov, A.~Chatrabhuti and H.~Isono,
``$\alpha$-attractors from supersymmetry breaking,''
Eur. Phys. J. C \textbf{81} (2021) no.2, 166
[arXiv:2009.02203 [hep-th]].

\bibitem{Co:2021ion}
R.~T.~Co, B.~Sheff and J.~D.~Wells,
``The Race to Find Split Higgsino Dark Matter,''
[arXiv:2105.12142 [hep-ph]].

\bibitem{Hall:2011jd}
L.~J.~Hall and Y.~Nomura,
``Spread Supersymmetry,''
JHEP \textbf{01} (2012), 082
[arXiv:1111.4519 [hep-ph]].
 
\bibitem{Tucker-Smith:2001myb}
D.~Tucker-Smith and N.~Weiner,
``Inelastic dark matter,''
Phys. Rev. D \textbf{64} (2001), 043502
[arXiv:hep-ph/0101138 [hep-ph]].

\bibitem{Tucker-Smith:2004mxa}
D.~Tucker-Smith and N.~Weiner,
``The Status of inelastic dark matter,''
Phys. Rev. D \textbf{72} (2005), 063509
[arXiv:hep-ph/0402065 [hep-ph]].
 
\bibitem{Nagata:2014wma}
N.~Nagata and S.~Shirai,
``Higgsino Dark Matter in High-Scale Supersymmetry,''
JHEP \textbf{01} (2015), 029
[arXiv:1410.4549 [hep-ph]].
 
\bibitem{Chung:1998rq}
D.~J.~H.~Chung, E.~W.~Kolb and A.~Riotto,
``Production of massive particles during reheating,''
Phys. Rev. D \textbf{60} (1999), 063504
[arXiv:hep-ph/9809453 [hep-ph]].
 
\bibitem{Berezinsky:2008bg}
V.~Berezinsky, M.~Kachelriess and M.~A.~Solberg,
``Supersymmetric superheavy dark matter,''
Phys. Rev. D \textbf{78} (2008), 123535
[arXiv:0810.3012 [hep-ph]].
 
\end{thebibliography}
\end{document}